# A Streamlit-based Artificial Intelligence Trust Platform for Next-Generation Wireless Networks

Murat Kuzlu[1], Ferhat Ozgur Catak[2], Salih Sarp[3], Umit Cali[4], Oezguer Gueler[5]
[1]*Electrical Engineering Technology, Old Dominion University*, Norfolk, VA, USA
[2]*Department of Electrical Engineering and Computer Science, University of Stavanger*, Rogaland, Norway
[3]*Electrical and Computer Engineering, Virginia Commonwealth University*, Richmond, VA, USA
[4]*Department of Electric Power Engineering, Norwegian University of Science and Technology*, Trondheim, Norway
[5] *eKare, Inc.,* Fairfax, VA, USA
mkuzlu@odu.edu, f.ozgur.catak@uis.no, sarps@vcu.edu, umit.cali@ntnu.no, oguler@ekareinc.com

*Abstract*—With the rapid development and integration of artificial intelligence (AI) methods in next-generation networks (NextG), AI algorithms have provided significant advantages for NextG in terms of frequency spectrum usage, bandwidth, latency, and security. A key feature of NextG is the integration of AI, i.e., self-learning architecture based on self-supervised algorithms, to improve the performance of the network. A secure AI-powered structure is also expected to protect NextG networks against cyber-attacks. However, AI itself may be attacked, i.e., model poisoning targeted by attackers, and it results in cybersecurity violations. This paper proposes an AI trust platform using Streamlit for NextG networks that allows researchers to evaluate, defend, certify, and verify their AI models and applications against adversarial threats of evasion, poisoning, extraction, and interference.

## I. INTRODUCTION

Next-generation networks (NextG or 5G and beyond) have become more popular as a new version of mobile communication in recent years after 1G-2G in the early 1990s, 3G in the early 2000s, and 4G in the 2010s. These networks can support high-speed data transmission with low latency for real-time data transmission, i.e., 5G offers a data transmission speed of 20 times faster than 4G networks and delivers less than a millisecond of data latency [1], [2]. The main difference of NextG is to use advanced communication methods, millimeter-wave massive multiple-input multiple-output (MIMO) [3], beamforming [4], [5], and intelligent reflection surfaces (IRS) [6]. 6G is one of the NextG networks following 5G, which promises 100 times faster mobile data speeds with lower latency than the 5G network, i.e., approximately 1 Tbps and 1 ms, respectively. In the near future, NextG will be used in the connectivity of cars, drones, mobile devices, the Internet of Things (IoT) devices, homes, and many more. The key component of NextG is the integration of AI, i.e., self-learning architecture based on self-supervised algorithms, to improve the performance of the network for tomorrow's cellular systems [7], [8]. Using AI algorithms provides novel solutions for massive MIMO systems involving a large number of antennas and beam arrays. For example, a beam codeword consists of analog phase-shifted values applied to the antenna elements to form an analog beam in [9], base beam selection with deep learning algorithms is proposed for using channel state information for the sub-6 GHz links. In the literature, most studies have focused on the communication methods to increase cellular technologies' performance but usually ignored the security and privacy issues and the integration of currently emerging AI tools into NextG. The study in [10] discusses key trends and AI-powered methodologies for 6G network design and optimization. A secure AI-powered structure is expected to protect NextG networks against cyber-attacks. However, AI itself may be attacked, i.e., the model poisoning by attackers resulting in cybersecurity violations. With the use of AI algorithms in NextG's physical layer functions, such as channel estimation, modulation recognition, and channel state information (CSI) feedback, the physical layer faces new challenges prone to adversarial attacks. The study in [11] indicates that an adversarial attack may have a destructive effect on DL-based CSI feedback, and transmitted data can be easily tampered with adversarial perturbation by malicious attackers due to the broadcast nature of wireless communication. These examples can be extended with other use cases using AI algorithms for NextG networks, e.g., the mmWave beam prediction for several base stations (BSs) with multiple users using deep learning algorithms. The integration of AI algorithms for 5G and beyond technologies can lead to potential security problems if not addressed properly. Mainly, most studies focus on building ML algorithms for NextG communication problems. However, there are limited studies focusing on the security threats against AI models used in NextG networks. The future of wireless communication will utilize more AI capabilities. The attack against AI models is different from well-known wireless physical layer security. The purpose of the attack on wireless physical layer security is to make the transmitted signal non-predictive to decrease the secrecy capacity. This paper proposes a Streamlit-based platform to address security concerns and improve AI trustworthiness in NextG networks. The developed platform allows researchers, engineers, and testers to evaluate and verify their AI models. The source code is available from GitHub [1].

[1]https://github.com/muratkuzlu/NextG_AI_Trust_Platform

## II. THE PROPOSED INTERACTIVE ARTIFICIAL INTELLIGENCE (AI) TRUST PLATFORM

The architecture of the proposed interactive Artificial Intelligence (AI) trust platform is shown in Fig. 1. The architecture includes the following steps: (1) Application Selection, (2) Load Data, (3) Load Model, (4) Fine-Tuning for Hyperparameters, (5) Attack Power, (6) Attack Model, (7) Model(s) Training and Evaluation, and (8) Applying Mitigation Method.

1) Application Selection: The framework will support beamforming, channel estimation, intelligent reflecting surface (IRS), and spectrum sensing. The proposed architecture in this paper allows only beamforming applications for now. Other applications will be added in the future.
2) Load Data: The first step is to acquire data from publicly available data resources or manual user upload. The platform reads data through the Streamlit server and loads the local server for further analysis. The platform allows uploading file(s) in CSV and MAT format.
3) Load Model: The platform allows users to load their AI-powered models and store them on a local server for prediction. The platform also hosts pre-trained models for next-generation network applications.
4) Fine-Tuning for Hyperparameters: The performance of AI-powered models significantly depends on model hyperparameters. A grid search method determines the best hyperparameters for each possible combination. However, it requires more computing processes. Hyperparameters are used to tune the model parameters, such as the learning rate, epoch and batch size, the optimizer, etc.
5) Attack Power Selection: The attack power decides the level of generated adversarial examples for AI-powered models. The platform offers four attack power levels: none, low, medium, and high. None means that an adversarial attack will not be applied to the model(s).
6) Applying Adversarial Attack: The platform hosts four widely used popular adversarial attack models, i.e., Fast Gradient Sign Method (FGSM), Basic Iterative Method (BIM), Projected Gradient Descent (PGD), Momentum Iterative Method (MIM).
7) Model(s) Training and Evaluation: The AI-powered models for next-generation network applications are trained and evaluated. The main objective of this step is to train the selected model(s) and to measure the model performance against adversarial attacks with and without adversarial training in terms of model accuracy, i.e., the Mean Average Error (MAE), the Mean Squared Error (MSE) and the Root Mean Squared Error (RMSE).
8) Applying Mitigation Method: The framework hosts adversarial training and defensive distillation mitigation methods. They are widely recommended defense techniques, which generate adversarial instances using the gradient of the victim classifier and then re-training the model with the adversarial instances.

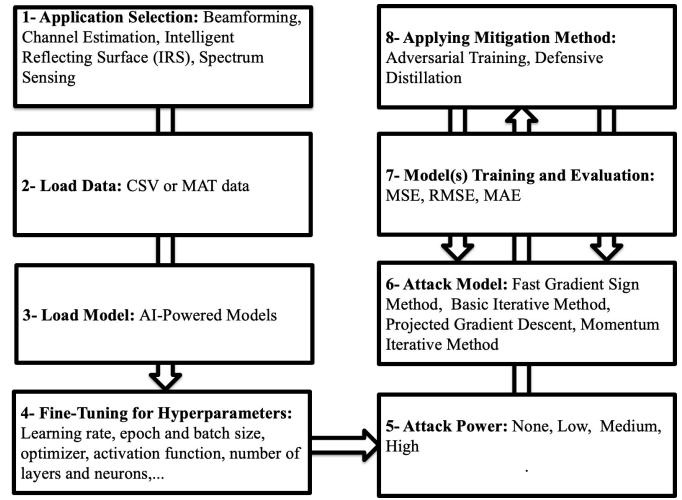

Fig. 1. The architecture of the proposed interactive Artificial Intelligence (AI) trust platform

The proposed platform focuses on AI-powered models for next-generation network applications and hosts four types of adversarial attacks (FGSM, BIM, PGD, and MIM) and two mitigation methods (adversarial training and defensive distillation) in the current version.

## III. ADVERSARIAL MACHINE LEARNING ATTACKS AND MITIGATION METHODS HOSTED BY PLATFORM

In this section, a brief overview of the adversarial machine learning attacks (FGSM, BIM, PGD, MIM), and mitigation methods (adversarial training and defensive distillation) hosted by the platform will be presented.

### A. Adversarial Machine Learning Attacks

Adversarial machine learning attacks generate adversarial samples close to real data to manipulate the trained model outputs. The attacker tries to generate a perturbation to the adversarial examples, which would affect the prediction phase of the machine learning model. Adversarial machine learning attacks work well if the attacker has access to the training data. Methods for constructing adversarial examples can be categorized into two groups: gradient-based and content-based attacks, respectively. In this study, gradient-based attacks were chosen as adversarial attacks because of their simplicity and variety. The adversarial attacks used in the proposed platform are briefly explained as follows:

- **Fast Gradient Sign Method (FGSM)**: The FGSM attack was proposed in [12] and the first attack using the gradient of the loss function. It was proposed using a one-step gradient-based method. This method consists of adding a small perturbation to the input example to manipulate the output of the machine learning model.
- **Basic Iterative Method (BIM)**: The BIM attack was proposed in [13] as an iterative extension of the FGSM attack. The BIM attack applies a small perturbation to the input example multiple times.

- **Projected Gradient Descent (PGD)**: The PGD attack was introduced in [14]. It is also an iterative extension of the FGSM and similar to the BIM. However, PGD starts the search for the adversarial example randomly, then runs iterations to find an adversarial example with the greatest loss with a smaller perturbation.
- **Momentum Iterative Method (MIM)**: The MIM attack was proposed in [15] as an extension of the BIM. This attack improves BIM's convergence to stabilize the gradient's direction at each step.

### B. Mitigation Methods

AI-powered models are vulnerable to adversarial machine learning attacks. Fortunately, the mitigation methods can improve the robustness of models. The mitigation methods used in this study, i.e., Adversarial training and defensive distillation, will be briefly explained.

*1) Adversarial Training:* Adversarial training is a widely used method to improve the robustness of a machine learning model. Adversarial training creates adversarial examples using the gradient of the victim classifier and then re-trains the model with the adversarial instances and their respective labels. The objective is to train the model with adversarial examples closer to real data to be less sensitive to perturbations. The following steps are used to apply for adversarial training. The steps for adversarial training are described as follows:
- Define the loss function: The loss function for a single training example is defined as $L(f, x, y) = I[y \neq f(x)]$. Here, $y$ is the ground truth label, and $f(x)$ is the predicted label. The objective of the loss function is to minimize the error in the prediction. The loss function for multiple training examples is defined as $L(f, X, Y) = \sum_{i=1}^{N} L(f, x_i, y_i)$. Here, $N$ is the number of training examples.
- Generate adversarial examples: The adversarial example is generated using the gradient of the loss function with respect to the input example. The adversarial example is defined as $x_{adv} = x + \epsilon \cdot \text{sign}(\nabla_x L(f, x, y))$. Here, $\epsilon$ is the perturbation magnitude, $x$ is the input example, and $\nabla L(f, x, y)$ is the gradient of the loss function with respect to the input example.
- Train the model: The model is trained on the original training examples as well as the generated adversarial examples. The objective function for the model is defined as $L(f, X, Y) = \sum_{i=1}^{N} L(f, x_i, y_i) + \alpha \sum_{i=1}^{N} L(f, x_i + \epsilon \cdot \text{sign}(\nabla_x L(f, x_i, y_i)), y_i)$. Here, $X$ is the training set, $Y$ is the labels, and $\alpha$ is the weighting parameter.

*2) Defensive Distillation:* Defensive distillation is another widely used method to improve the robustness of a machine learning model. There are two models involved in defensive distillation. The first model is the teacher model, which is used to train the second model, i.e., the student model. The teacher model is trained using the original training data, and the student model is trained using the output of the teacher model. Defensive distillation aims to train the student model to be less sensitive to perturbations. The following steps are used to apply for defensive distillation. The steps for defensive distillation are described as follows:
1) Train the teacher model: The teacher model is trained using the original training data.
2) Generate the soft probabilities: The output of the teacher model can be used to generate the soft probabilities. The soft probability is defined as $q_i = \text{softmax}(z_i)$, where $z_i$ is the output of the $i$-th neuron.
3) Train the student model: The student model is trained using the output of the teacher model, i.e., the soft probabilities. The objective function for the student model is defined as $L(f, X, Y) = \sum_{i=1}^{N} L(f, q_i, y_i)$. Here, $X$ is the training set, $Y$ is the labels, and $q_i$ is the output of the $i$-th neuron.

## IV. CASE STUDY AND DEMONSTRATION

### A. Dataset Description and Case Study

Streamlit-based AI trust platform evaluates the security of beamforming AI models. The beamforming is the radio frequency (RF) management, which provides wireless signal connection towards specific receivers for faster and more reliable communication. It also reduces interfering signals with more focused beamed signals instead of omnidirectional broadcasting. Many factors affect the beamforming prediction, such as the users' locations, BSs, and any obstacles. A DL model is developed to study the reflections and diffractions of the pilot signal generated using DeepMIMO-based beam patterns. The proposed DL-based beamforming prediction comprises two states, i.e., training and prediction.

In the training stage, pilot signals are used to outline the estimation of channels. The uplink training pilot sequences are sent from the user to the BSs for each beam coherence time. BSs responsibility is to merge training pilot sequences on RF beamforming vector. Each pilot sequence of uplink data is synchronously sent to BSs to be used for channel estimation and precoding. These sequences are then utilized as inputs for the DL model to estimate the achievable rate for each RF beamforming vector.

The RF beamforming vectors are estimated using the trained DL-based model in the prediction phase. Similar to the training phase, each user sends an uplink pilot sequence to the BSs, where all coming sequences are combined and sent to the cloud. These sequences are fed to the DL model to predict the best beamforming vectors using maximum achievable rates for each BS. Then, predicted beamforming vectors are used to estimate the effective channel.

### B. Demonstration

AI trust platform dashboard consists of two parts: (1) Settings and (2) Experimental results. The settings part includes all parameters from load model, load data, fine-tuning, attack power selection, to attack model and mitigation model selection, as shown in Figure 2 (left side). This section demonstrates the visualization of experimental results based on the settings. The pre-trained model and input/output dataset are loaded, and all settings are selected by the user before running the experiment. In this experiment, the beamforming application is selected along with all possible attack power levels (All), the attack model (FGSM), and the mitigation method (Adversarial Training). Experiential results are shown in Figure 2 (right side). The results are given as a table and a figure. The platform also allows exporting results as a CSV file. According to the results, the AI-powered beamforming model used in NextG networks is vulnerable to adversarial attacks. The model becomes more vulnerable to high attack power. On the other hand, FGSM is not a powerful attack on the model due to its simplicity. Fortunately, the mitigation methods, i.e., defended models, can help the improvement of the model's robustness under FGSM attacks. The figure shows that MSE values are much lower than the undefended results, i.e., without a mitigation method.

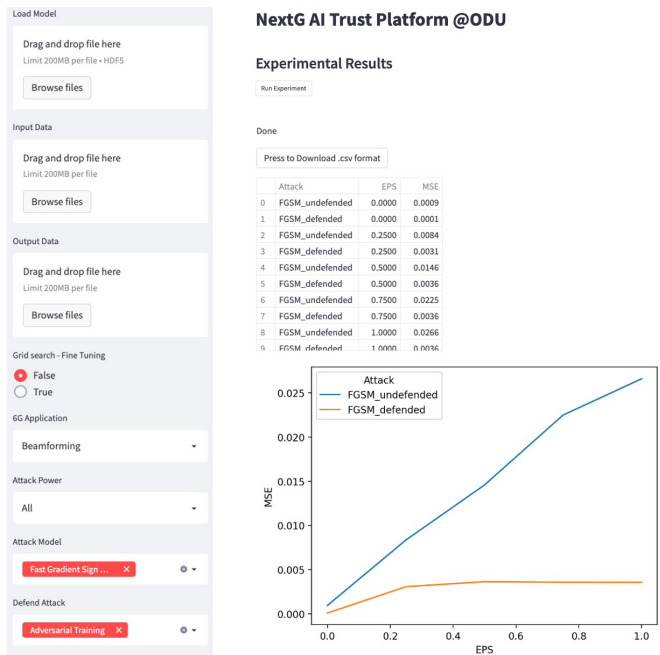

Fig. 2. The demonstration of the experimental results

## V. CONCLUSION

This paper proposes a Streamlit-based AI trust platform for NextG networks focusing on beamforming applications to improve AI trustworthiness. The platform hosts well-known adversarial machine learning attacks, such as FGSM, BIM, PGD, and MIM, along with the mitigation methods, i.e., adversarial training and defensive distillation. It is expected that researchers and engineers can test and evaluate their models in terms of performance and robustness against adversarial attacks. By using Streamlit library, the process of creating a web-based platform for any AI model makes it flexible and interactive. A user can load data and models through the platform and select any adversarial attack, mitigation method, and attack power level. The platform dashboard shows the results, i.e., the model's vulnerability, based on the user selection. The authors are planning to extend the platform to other AI-powered NextG applications, such as channel estimation, spectrum sensing, and IRS, and make it more flexible and interactive.

ACKNOWLEDGMENT

This work was supported in part by the Commonwealth Cyber Initiative, an investment in the advancement of cyber R&D, innovation, and workforce development in Virginia.

## REFERENCES

[1] H. Viswanathan and P. E. Mogensen, "Communications in the 6g era," *IEEE Access*, vol. 8, pp. 57 063–57 074, 2020.
[2] E. C¸ atak and L. Durak-Ata, "Waveform design considerations for 5g wireless networks," *Towards 5G Wireless Networks-A Physical Layer Perspective*, pp. 27–48, 2016.
[3] Z. Pi and F. Khan, "A millimeter-wave massive mimo system for next generation mobile broadband," in *2012 Conference Record of the Forty Sixth Asilomar Conference on Signals, Systems and Computers (ASILOMAR)*. IEEE, 2012, pp. 693–698.
[4] F. O. Catak, M. Kuzlu, E. Catak, U. Cali, and D. Unal, "Security concerns on machine learning solutions for 6g networks in mmwave beam prediction," *Physical Communication*, p. 101626, 2022.
[5] M. Kuzlu, F. O. Catak, U. Cali, E. Catak, and O. Guler, "The adversarial security mitigations of mmwave beamforming prediction models using defensive distillation and adversarial retraining," *arXiv preprint arXiv:2202.08185*, 2022.
[6] S. Gong, X. Lu, D. T. Hoang, D. Niyato, L. Shu, D. I. Kim, and Y.-C. Liang, "Toward smart wireless communications via intelligent reflecting surfaces: A contemporary survey," *IEEE Communications Surveys & Tutorials*, vol. 22, no. 4, pp. 2283–2314, 2020.
[7] Y. Xiao, G. Shi, Y. Li, W. Saad, and H. V. Poor, "Toward self-learning edge intelligence in 6g," *IEEE Communications Magazine*, vol. 58, no. 12, pp. 34–40, 2020.
[8] T. Huang, W. Yang, J. Wu, J. Ma, X. Zhang, and D. Zhang, "A survey on green 6g network: Architecture and technologies," *IEEE access*, vol. 7, pp. 175 758–175 768, 2019.
[9] M. S. Sim, Y.-G. Lim, S. H. Park, L. Dai, and C.-B. Chae, "Deep learning-based mmwave beam selection for 5g nr/6g with sub-6 ghz channel information: Algorithms and prototype validation," *IEEE Access*, vol. 8, pp. 51 634–51 646, 2020.
[10] C. Yizhan, W. Zhong, H. Da, and L. Ruosen, "6g is coming: Discussion on key candidate technologies and application scenarios," in *2020 International Conference on Computer Communication and Network Security (CCNS)*. IEEE, 2020, pp. 59–62.
[11] Q. Liu, J. Guo, C.-K. Wen, and S. Jin, "Adversarial attack on dl-based massive mimo csi feedback," *Journal of Communications and Networks*, vol. 22, no. 3, pp. 230–235, 2020.
[12] I. J. Goodfellow, J. Shlens, and C. Szegedy, "Explaining and Harnessing Adversarial Examples," *arXiv e-prints*, p. arXiv:1412.6572, Dec. 2014.
[13] A. Kurakin, I. Goodfellow, and S. Bengio, "Adversarial examples in the physical world," *arXiv e-prints*, p. arXiv:1607.02533, Jul. 2016.
[14] A. Madry, A. Makelov, L. Schmidt, D. Tsipras, and A. Vladu, "Towards Deep Learning Models Resistant to Adversarial Attacks," *arXiv e-prints*, p. arXiv:1706.06083, Jun. 2017.
[15] Y. Dong, F. Liao, T. Pang, H. Su, J. Zhu, X. Hu, and J. Li, "Boosting Adversarial Attacks with Momentum," *arXiv e-prints*, p. arXiv:1710.06081, Oct. 2017.